\input{aipcheck}
\documentclass[
   ,final            
 ]
 {aipproc}

\layoutstyle{6x9}

\begin{document}

\title{Variability type classification of multi-epoch surveys}

\classification{95.10.Gi, 95.75.De, 95.75.Fg, 95.75.Pq, 95.75.Wx, 95.85.Kr, 97.10.Jb, 97.10.Kc, 97.10.Sj, 97.10.Vm, 97.10.Yp, 97.10.Zr, 97.30.-b, 97.80.-d, 98.54.-h}
\keywords{Quasars, Variable stars}

\author{L. Eyer}{
 address={Geneva Observatory, Geneva University, CH-1290 Sauverny, Switzerland}
}

\author{A. Jan}{
 address={Geneva Observatory, Geneva University, CH-1290 Sauverny, Switzerland}
}

\author{P. Dubath}{
 address={Geneva Observatory, Geneva University, CH-1290 Sauverny, Switzerland}
}

\author{K. Nienartowicz}{
 address={Geneva Observatory, Geneva University, CH-1290 Sauverny, Switzerland}
}
\author{J.Blomme}{
 address={KUL, Leuven, Belgium}
}
\author{J.Debosscher}{
 address={KUL, Leuven, Belgium}
}
\author{J.De Ridder}{
 address={KUL, Leuven, Belgium}
}

\author{M. Lopez}{
 address={SVO / LAEFF-CAB (INTA-CSIC), LAEFF-ESAC, 28691 Villanueva de la Ca\~{n}ada, Madrid, Spain}
}

\author{L. Sarro}{
 address={Dpt. de Inteligencia Artificial, UNED, 28040 Madrid, Spain}
}

\begin{abstract}
The classification of time series from photometric large scale surveys into variability types and the description of their properties is difficult for various reasons including but not limited to the irregular sampling,  the usually few available photometric bands, and  the diversity of variable objects. Furthermore, it can be seen that different physical processes may sometimes produce similar behavior which may end up to be represented as same models. In this article  we will also be presenting our approach for  processing the data resulting from the Gaia space mission. The approach may be classified into following three broader categories: supervised classification, unsupervised classifications,  and "so-called" extractor methods i.e. algorithms that are specialized for particular type of sources. The whole process of classification- from classification attribute extraction to actual classification- is done in an automated manner. 
\end{abstract}

\maketitle

\section{Variability physical types and variability behaviours}

The classification of variable objects is necessarily based on the observable attributes. The lightcurve behaviour is the main parameter but some other source characteristics -such as mean absolute luminosity and mean colour- are equally useful. The ultimate goal is however to try to separate sources into physical
categories as a first step towards learning more about the nature of the source
and the causes for variability. Reaching this goal is complicated
by the fact that some very different physical processes can generate
similar variability behaviours. For example, two pulsating stars may end up having different models
to describe them, while two different physical process may be
described by the same models e.g. {\small EW} eclipsing binaries and some
mono-periodic pulsating stars.

The general-purpose classification may then be confused being unable
to disentangle two or more physical categories. Obviously a refined analysis may unravel the physical process, such as the analysis of luminosity, temperature, behaviours at different wavelengths etc. Therefore we have introduced a {\it
 specific object studies} module, whose goal is to validate and refine the
results of the general-purpose classification process.

Eyer and Mowlavi \cite{EyerMowlavi2007} proposed a tentative
organisation of variability physical types into a tree structure.  The
variability behaviours on the other hand, can be at the high-level,
be divided into different categories i.e. periodic, semi-regular,
irregular, and transient. The time scale and amplitude of the variation
can be very different: from event as short as a few seconds to secular
evolution and from milli-magnitudes to several magnitudes.

\section{Survey properties}
Different surveys have different properties such as photometric bands, random and systematic photometric errors, time sampling, possible crowding issues, etc. These specific properties can be exploited as well as possible through the data analysis in order to learn more about the physical processes at the origin of the variability. Here we consider the time sampling properties only.

There are two important ways to assess the quality of the time sampling: the time-lag histogram and the spectral window.

A {\bf time-lag histogram} is built from all possible time differences ($t_i-t_j$ such that $i > j$) between pair of measurements of a given lightcurve.  The time-lag histogram shows which delta times, and hence which variability timescales, are probed by a set of measurements. This is most useful to predict the ability of a survey to characterise transient or irregular variability behaviour of a given characteristic variation time.

The {\bf spectral window} is the Fourier transform of the set of sampling times. There are two extreme cases. If the measurements are taken at regular time interval, the Fourier transform will exhibit  series of strong peaks spaced by the frequency corresponding to this time interval. If, on the other hand, the time sampling is random, there will be no peak in the spectral window but a rather smooth continum. The Fourier transform of the signal is the convolution of the source signal with the spectral window. A mono-periodic source is a single spike in the Fourier domain, and in this case the Fourier transform of the signal has the shape of the spectral windows, but it is just shifted in frequency by the source frequency value. This allows to confirm our intuition: a random sampling will be much better for detecting periods. Note that period much smaller that the smallest time lag can in principle be recovered if the signal accuracy is high and the spectral window good \cite{EyerBartholdi1999}.

\begin{figure}
\includegraphics[height=.4\textheight]{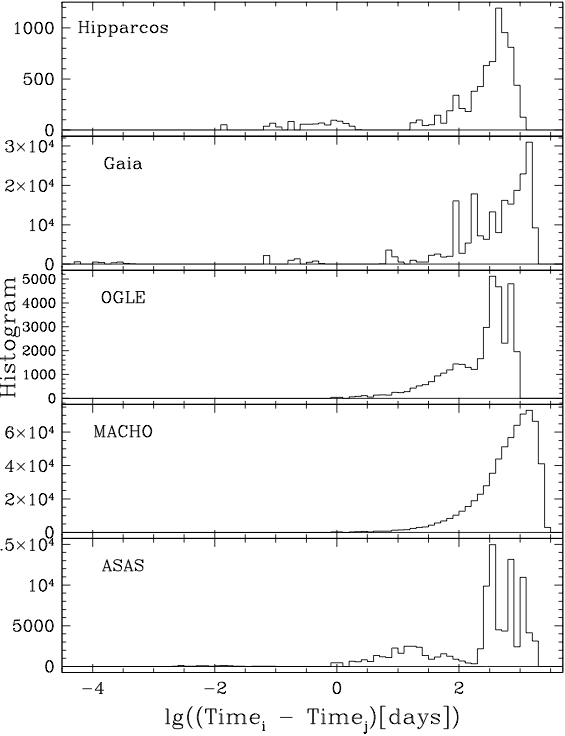}
\includegraphics[height=.4\textheight]{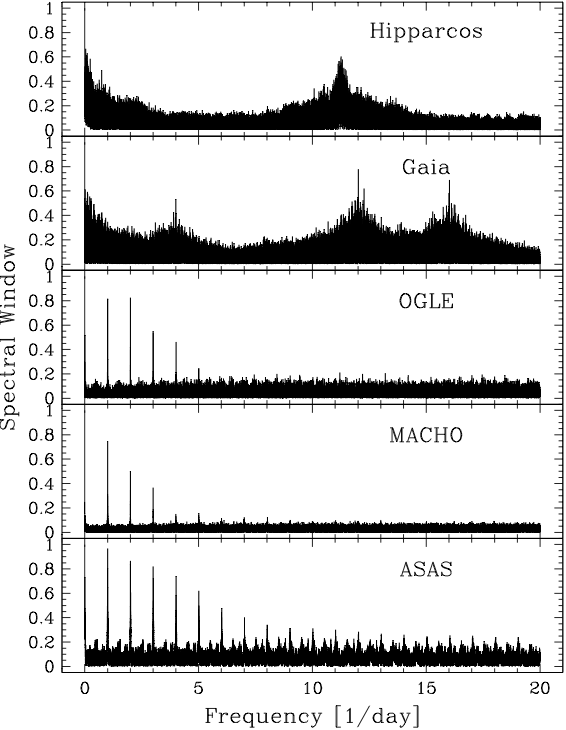}
\caption{We present two characteristics representing surveys: left: Histogram of time differences of different surveys ; right: Spectral window of different surveys. The time lag is given for Gaia on the per CCD photometry.\label{fig:sampling}}
\end{figure}

Both the time-lag histogram and the spectral window from several surveys (Hipparcos, Gaia, OGLE, MACHO and ASAS) are represented in Fig.~\ref{fig:sampling}.  For each case a random time series has been used to produce the diagrams. These surveys appear to have very different properties. Hipparcos and Gaia have similar spectral windows. They are quite good for period search as they show no strong secondary peaks. The global "noise" level is however higher because of the smaller number of measurements. In the other single site ground-based surveys, we see the peaks produced by the 1/day observation frequency, strongest in the case of ASAS. These figure allow us to better understand why Hipparcos was so successful in discovering complex objects like $\gamma\,$Dor and SPB stars with periods between 1-3 days.

\section{Some examples of classification methods applied to surveys}
Many studies based on microlensing searches focused on a given variable type and "extract" objects of this particular type using a priori knowledge about their variability properties. Variability types are extracted and studied independently and sequentially. Such a scheme is followed in most OGLE or MACHO studies. Others approaches based on global classification algorithms have started to flourish during this past decade. The goal of this section is not to go through all classification methods that appeared in the literature, but rather to show the diversity of approaches, using different sets of attributes or classification methods. To our knowledge, \cite{Waelkensetal1998} presented the first analysis which tackled a fully automated  classification on a whole sky survey. It was however limited to a rather small sample of B stars.  Using the combined information of characterisation of Hipparcos time series (such as main  period, amplitude in Hp band) and Geneva multicolour photometry, and  applying a multivariate discriminant analysis, the Hipparcos B stars  were classified into $\beta$Cep, SPB and Chemically Peculiar stars, and  also as eclipsing binaries. As far as classification attributes are concerned, \cite{EvansBelokurov2004} proposed to work in the Fourier space by defining an envelope of the power spectrum of the signal. Wyrzykowski et al. \cite{Wyrzykowskietal2003},  \cite{Wyrzykowskietal2004} used a neural network on the images of the folded time series. In the case of ASAS, Pojmanski and Maciejewski
\cite{PojmanskiMaciejewski2004} have set up an automated pipeline. First they separate the stricly periodic objects with respect to the
less regular ones and then they use the attributes from the ASAS light curve and 2MASS colours to perform the classification using the
variability types properties in ad-hoc selected projected 2D planes. Eyer \& Blake \cite{EyerBlake2002}, \cite{EyerBlake2005} used an unsupervised classification method (Autoclass) applied on a Fourier modelling of the light curves of ASAS data.

\section{Plan for the Gaia mission}
CU7 is the Coordination Unit, part of the Gaia ground-based data processing and analysis consortium (cf. \cite{Mignardetal2008}), in charge of all aspects of the analysis of the measurement of variability. One of the most important tasks in this context is the classification of the sources according to their lightcurve behaviour. The classification is structured into a three step process (1) a number of attributes are first computed to characterise the lightcurves, (2) these attributes are then fed into the classification algorithms, and (3) finally specific processing is applied to the sources of each of the different groups obtained through step 2 to validate and possibly refine the classification result.
The different types of classification algorithms considered are categorised into (1) supervised, (2) unsupervised, and (3) extractor
methods.

{\bf Supervised classification.} In the frame of supervised methods, the most important factor is that of finding best attributes in order to build the training set for classification algorithms. The strategy is to iteratively refine the training set by using different attribute/method combinations, then comparing the result. The most meaningful attribute set, both in terms of separation power and independence, can thus gradually be derived.

Although the training set data may be derived from the theoretical model,
the current approach is to rely on the data from existing surveys and by utilizing the work of \cite{Debosscheretal2007} done in the context of the
COROT mission. Following two steps are identified in order to build a representative training set for the Gaia variability classification task: firstly, a list of sources representing different variability class is to be established; secondly, lighcurves, similar to the ones obtained by the Gaia
mission, must be collected for selected source classes from the existing
data. For the second step, we will eventually be able to use Gaia
measurements during the mission as well. The final training set is likely to
emerge  from the iterative process as described above. In this regards, a multi-stage approach design by \cite{Sarroetal2008}, is being explored. The multi stage classifier breaks the classification problem into various stages, each classifying a specific set of source class possibly using different attribute sets. The main advantage of this approach is that the attribute sets and the classification schemes used at various stages could possibly be completely different for each stage

{\bf Unsupervised classification.} The challenge of the unsupervised classification with the Gaia data may well come
from the potentially large number of variable sources most probably of the order of
$10^8$. One of the first tasks is  to evaluate as to how many
sources can be fed into existing algorithms with currently available hardware
and then to try to extrapolate the results to the hardware that will eventually be
available during the mission. We may need to investigate algorithms that offer support for processing data in sub-samples in one way or another
before aggregating  the results. We also consider new time series indexing techniques to obtain intrinsic clustering directly in our data storage system with possibility of exploring novel data mining approaches in  very large databases. This may reduce the need for linear increase in demand for RAM, needed in a brute force approach.

{\bf Extractors.} Extractors are tools that take advantage of the knowledge gained about the
lightcurve behaviours for certain types of variable stars. Scanning the
complete set of variable objects, they try to identify specific
lightcurve behaviours, and thus they can ``extract'' sources of a
given class. The current list of extractors that CU7 plan to develop
includes: (1) Microlensing events, (2) Flare Stars.

{\bf Specific Object Studies (SOS).} Once the sources have been classified by general algorithm into
classes, we plan to carry out a specific processing for each of the
classes. Two important goals of the SOS processing is to validate and
refine the results of the classification.

\subsection{Classification of Hipparcos sources}
\begin{figure}
\includegraphics[height=.30\textheight]{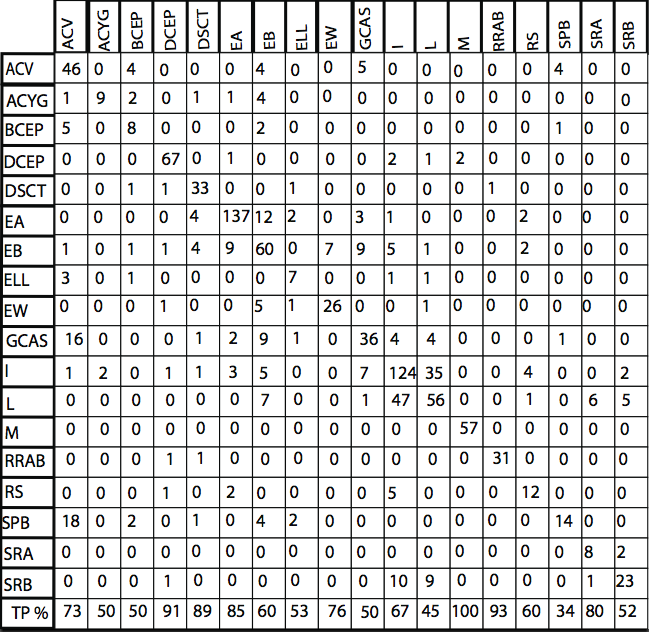}
\caption{The confusion matrix resulting from classification process\label{fig:confusionmatrix}}
\end{figure}

Of all the available surveys, Hipparcos shares most of the aspects of the Gaia mission i.e. the global sky survey, astrometric information, its peculiar sampling etc. Furthermore, till date the whole Hipparcos data has not been classified in a systematic way; the previous studies \cite{ESA1997}, \cite{Waelkensetal1998} and \cite{Aertsetal1998} either employed visual inspection methods for classification or investigated objects from specific (sub)classes. \cite{WillemsenEyer2007} presented the results of first ever systematic global classification of the Hipparcos data, using support vector machine, and formed basis for the current study. We employed a two approaches in order to undertake the classification task. In one approach, we utilized the attribute sets as established by  \cite{WillemsenEyer2007} and in other approach the characterization module, part of CU7 software framework, was used to extract key attributes from the Hipparcos light curves. Some of the important attributes include standard deviation (weighted), skewness (weighted), kurtosis (weighted), normalized p2p scatter (weighted), slope, log frequency, amplitude, phase, log amplitude, range and log range, B-V and V-I. The classification models were obtained using support vector machine algorithm being part of Weka machine learning library, and were applied using CU7 classification module. It may be noted that using the current implementation of CU7 software framework, the complete process of extracting classification attributes, and performing classification can be done in a completely automated manner. We used different strategies to validate the classifier result i.e. using n-fold cross validation as well as doing a percentage split of the available data in the training and test sets. Using a training set of 2100 objects and a test set of 1100 objects, a classification accuracy of nearly 68 percent was achieved, with more than 9 classes having an accuracy well above 70 percent (i.e. above 90 percent in some cases).  The following table summarizes the confusion matrix for the classifier obtained using the attributes from \cite{WillemsenEyer2007}, the confusion matrix obtained using CU7 characterization module was also similar to the one displayed in Fig~\ref{fig:confusionmatrix}.

\bibliographystyle{aipproc}   

\bibliography{Eyeretal}

\IfFileExists{\jobname.bbl}{}
{\typeout{}
 \typeout{******************************************}
 \typeout{** Please run "bibtex \jobname" to optain}
 \typeout{** the bibliography and then re-run LaTeX}
 \typeout{** twice to fix the references!}
 \typeout{******************************************}
 \typeout{}
}

\begin{theacknowledgments}
We warmly thank M.Beck, L.Guy, I.Lecoeur and N.Mowlavi, as well as the CU7 members who are contributing to the pipeline.
\end{theacknowledgments}

\end{document}